# High-Temperature Hydrogen Sensors Based on Gallium Oxide Heterojunction Diodes

William A. Callahan[1,*], Kingsley Egbo[1], Anna Sacchi[1], Michelle Smeaton[1], Michael Walker[2], Anna Staerz[2], Ryan O'Hayre[2], Andriy Zakutayev[1,*]

[1]Materials Science Center, National Laboratory of the Rockies

[2]Department of Metallurgical and Materials Engineering, Colorado School of Mines

* Contact: will.callahan@nlr.gov, andriy.zakutayev@nlr.gov

**Abstract:**

Long-term, high temperature operation of $Ga_2O_3$ devices is a crucial hurdle that must be overcome before widespread adoption of the technology can be achieved, but is largely absent from the overall body of work. Demonstrations up to this point show devices are either limited by material or dopant instability that leads to performance degradation with time. Herein, $Ga_2O_3$-based hydrogen sensors employing Pt Schottky and $Cr_2O_3/Ga_2O_3$ p-n diodes (Mg- and N-doped) were fabricated and evaluated for long-term stability at 600°C for 800–1,800 hours, with cyclic exposure to $N_2$ and low-concentration $H_2$ (500–1,500 ppm). Transient current density (measured at -0.1 V) and periodic J-V characterization were used to track performance. Despite gradual declines in sensor signal and sensitivity, devices distinguished hydrogen concentrations throughout weeks of operation. Degradation was architecture-dependent: $Cr_2O_3$:Mg degraded gradually, consistent with known Mg migration; the Pt Schottky diode showed dramatic changes after 1,000 hours; and $Cr_2O_3$:N showed the lowest but most stable performance before failing at 800 hours. Thermionic emission and Lambert W-based modeling confirmed hydrogen exposure reduces interfacial barrier height via a proton-induced dipole mechanism common to both diode types. TEM of aged Pt Schottky diodes revealed Pt grain growth and microvoid formation as key degradation mechanisms. TOF-SIMS confirmed nitrogen dopants remain confined to the $Cr_2O_3$:N layer, supporting N-doping as a stable, lower-performance alternative to Mg-doping.

## 1. Introduction:

Reliable electronic device operation in harsh, high-temperature environments is critical for industrial, transportation, and energy applications, with gas sensors serving a key early-warning function. Among the most fundamental is the hydrogen gas sensor. Hydrogen's flammability limit drops significantly at elevated temperatures and under oxygen-poor conditions (1–4), making sensitive detection essential for safety, particularly given industry standards demanding fast response, wide operating range, long lifetime, and minimal cross-sensitivity (5). Hydrogen is central to several high-temperature power technologies, including solid oxide fuel cells (600–1000°C) (6–8); hydrogen-fueled gas turbines operating in hypoxic exhaust streams (250–850°C ) (7–10); and down-hole geothermal, oil, and gas applications, where temperatures can reach up to 500°C in extreme reservoirs (11–17); all of which require robust, high-temperature-compatible sensing solutions.

Existing commercial hydrogen-sensing technologies each face limitations at high temperature or under hypoxic conditions: semiconductor metal oxide sensors (e.g., $SnO_2$, $WO_3$) require oxygen to function (18–24); surface acoustic wave sensors lose piezoelectric response above 550°C (25–27); and electrochemical YSZ-based sensors are restricted to a relatively narrow 400–600°C operating window (28–31). Diode-based sensors offer an alternative, using catalytic metals (Pt,

Pd, or alloys) to dissociate $H_2$ and increase reverse-bias leakage current via a thermionic emission mechanism (32,33), though Schottky diodes are limited by modest barrier heights and thermal saturation of reverse current at high temperature (34,35). Alternatively, p-n diodes, while largely unexplored for hydrogen sensing, offer a similar sensing mechanism with larger barrier heights and lower leakage current, potentially avoiding thermal saturation issues (36).

Beta gallium oxide ($\beta$-$Ga_2O_3$) has received significant attention in the last decade as a candidate material for high-temperature and high-power applications. Due to its large bandgap, thermal excitation of intrinsic carriers and the resulting increase in leakage current occurs at significantly higher temperatures than materials like silicon, while retaining the ability for lower-cost, bulk crystal growth techniques like EFG and the Czochralski method.

The most common studies of high-temperature operation gallium oxide device performance have examined a single set of temperature exposure curve (37–40). Fewer in number are those that study the change performance at each temperature after temperature cycling (≤ 10 cycles) (41) or longer-term continuous operation at a higher temperature (42). Our previous studies have aimed to expand upon that body of work through systematic re-examination of various contact structures, starting with developing a stable Ohmic contact (43) and most recently demonstrating a novel $Cr_2O_3$:Mg-based pn heterojunction diode that operated continuously at 600˚C for more than 160 hours (44). Both of these studies incorporated long-term holds and repeated thermal cycles to gauge thermochemical and thermomechanical stability of the systems.

This work extends prior efforts to improve the high-temperature reliability of $Ga_2O_3$-based devices by repeatedly exposing three diode architectures — Pt/β-$Ga_2O_3$ (Schottky), Pt/$Cr_2O_3$:Mg/β-$Ga_2O_3$ (p-n), and Pt/$Cr_2O_3$:N/β-$Ga_2O_3$ (p-n) — to low hydrogen concentrations at 600˚C for extended durations (800–1800 hours, depending on device), using potentiostatic measurements at -0.1 V to simulate "always-on" operation, supplemented by twice-daily full J-V sweeps. Results show that while performance degraded steadily over time, sensitivity to all three $H_2$ concentrations persisted throughout testing, and diode model parameter extraction confirms that hydrogen exposure lowers the potential barrier height across all device types. Notably, prior work identified Mg diffusion to the $Cr_2O_3$:Mg/$Ga_2O_3$ interface, and its subsequent oxidation into an insulating MgO layer, as a key driver of performance decline via increased stack resistance, motivating the comparison with the N-doped alternative. Overall, this work establishes a foundational device architecture for further optimization through improved materials, growth strategies, and encapsulation.

## 2. Methods:

### *2.1 Device fabrication*

Device fabrication was completed using the following process: as-received HVPE (001) β-$Ga_2O_3$ substrates (Novel Crystal Technologies) were cleaned via an organic wash, SPR etch, and DI water rinse. Large-area Ohmic back contacts (5 nm Ti/100 nm Au) were then deposited on the substrate backside by e-beam evaporation using a Temescal FC2000 system under high-vacuum conditions, with all metal layers deposited sequentially without breaking vacuum, followed by annealing at 550°C for 90 s in flowing $N_2$.

Following the back-contact anneal, one of two routes was used to form the p-type layer for p-n diode devices: (1) Mg-doped $Cr_2O_3$ ($Cr_2O_3$:Mg) was deposited via pulsed laser deposition (PLD) using an 8% Mg-doped $Cr_2O_3$ target, or (2) nitrogen-doped $Cr_2O_3$ ($Cr_2O_3$:$N_2$) was deposited using an undoped $Cr_2O_3$ target in flowing $N_2$. A 30 nm Pt contact was then deposited either directly on a $Ga_2O_3$ substrate to form a Schottky diode, or on top of the newly formed p-type layer to form an Ohmic contact for the $Cr_2O_3$ p-n diodes. Shadow masks used to define a 2×6 grid of 1 mm diameter circular contacts. This platinum recipe was formulated to increase adhesion to the polished $Ga_2O_3$ surface (45–47) and was programmed with an especially slow deposition rate for the first few nanometers. Details about the slow Pt recipe can be found in the Supplemental Information. Platinum was selected for this metallization step due to its strong catalytic activity for hydrogen dissociation, its thermodynamic stability, and its favorable work function alignment to both semiconductor materials. A schematic of the architecture and a picture of a fabricated device are shown in Figure 1. More details about deposition rationale and techniques can be found in previous reports (43,44).

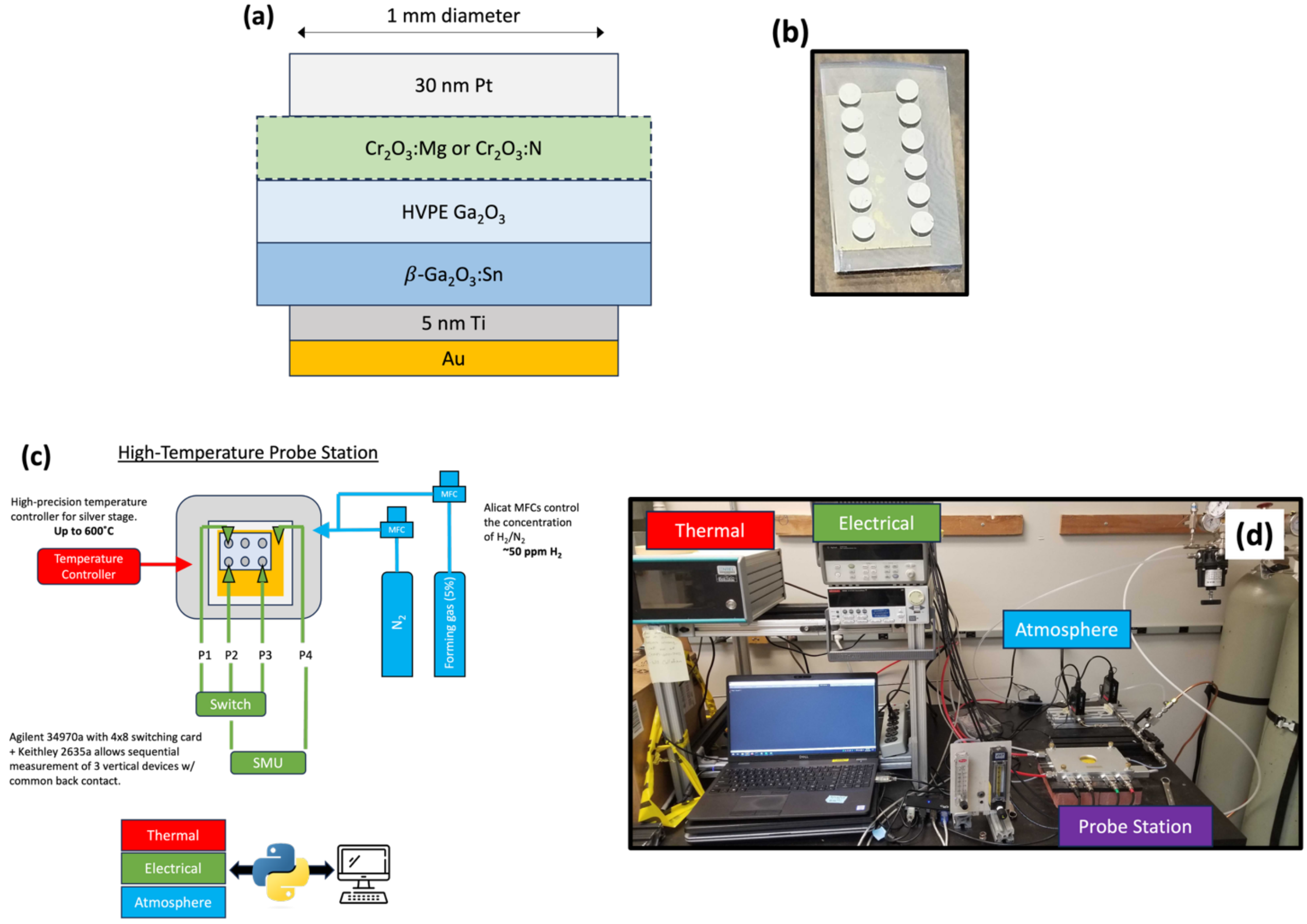

Figure 1: (a) (Left) Device architecture for the vertical diodes used in this study. (Right) Picture of a fabricated device on a 5 mm × 10 mm substrate. Platinum electrodes are deposited in a 2 × 6 array of 1 mm diameter circles, as defined by shadow mask. (b) Instrument schematic detailing the thermal, atmospheric, and electrical control elements of the measurements. (c) schematic and (d) photo of experimental setup.

Each chip consists of twelve 1-mm diameter vertical diodes sharing a large-area back contact, formed by depositing a 5 nm Ti/100 nm Au metal stack onto a thin alumina substrate using the previously described evaporation system; this back platform serves both as a handle for the chip and as a landing pad for the SMU sink contact, with a small drop of silver paint (Alfa Aesar Leitsilber) used to improve thermal and electrical contact between the alumina platform and the Ga2O3 chip. Measurements were performed using an Instec HCP621G-PMH high-temperature probe station with a Keithley 2635A SMU and an Agilent 34970A switching unit to sequentially measure three devices via their shared back contact. Gas delivery and mixing were controlled by a pair of Alicat mass flow controllers, with one gas line flowing $N_2$ and the other flowing forming gas (5% $H_2$, balance $N_2$), producing nominal $H_2$ setpoints of based on the proportional mixing given a known total flow rate.

### *2.2 Long-term electrical measurements*

Prior to the long-term device testing, a 'staircase' measurement was performed to characterize the device response to $H_2$. This measurement was performed as a function of both temperature (100–600°C in 100°C steps) and hydrogen concentration (0, 1250, 2500, 5000, and 12500 ppm nominal, balanced with $N_2$ at a total flow rate of 200 sccm). A dedicated, unconditioned device with no prior high-temperature exposure was used for this experiment. At each temperature, the device was allowed to equilibrate for 30 minutes in $N_2$, which also served as a blanking step, followed by 15-minute exposures to each hydrogen concentration; the current at -0.1 V was recorded continuously throughout, without interspersed J-V curves, to maintain continuity at the sensing voltage.

The electrical performance of each of the three fabricated hydrogen sensors was separately and continuously monitored over a thermal soak exceeding 800, 1000, and 1800 hours at 600°C for the $Cr_2O_3$:N, $Cr_2O_3$:Mg, and Pt Schottky diodes respectively , with alternating exposure to low hydrogen concentrations followed by nitrogen purging. Characterization was performed using an Instec HCP621G-PMH probe station in combination with a Keithley 2635A SMU and an Agilent 34970A switching unit, with blended gas delivery controlled by two Alicat mass flow controllers supplying ultra-high-purity (UHP) $N_2$ and 5% $H_2$ forming gas (balance $N_2$), shown in Figure 1c and 1d.

Long-term electrical characterization consisted of two complementary measurement types. The first, referred to as the "hold" measurement, constantly recorded the current density of a single diode held at a constant -0.1 V. A single gas cycle consisted of an 11-hour $N_2$ segment to establish a baseline signal, followed by sequential exposure to three hydrogen concentrations (500, 1,000, and 1,500 ppm), each maintained for one hour. These hydrogen exposure steps were fit to a logistic growth function to more accurately extract the plateau value; more information can be found in the Supplemental Information. The second type of measurement consists of current-density-voltage measurements at the end of the nitrogen baseline portion and the end of the 1,500 ppm $H_2$ exposure, collected sequentially for three different diodes on the same chip via their shared back contact. As a result. J-V curves were measured for diodes on each chip at two points in each gas cycle (in $N_2$ and in 1,500 ppm $H_2$), approximately twice per day. This measurement scheme enabled simultaneous tracking of a single diode's transient behavior, as well as a more complete view of device health via J-V curves, which also provided data for diode parameter extraction and modeling. This scheme offers insight into the underlying sensing mechanism and allowing the performance evolution of the sensors to be assessed over time. This full routine was repeated for the duration of the thermal soak. Each portion of the hydrogen exposure was fit to a logistic function.

### *2.3 Material characterization*

Time-of-flight secondary ion mass spectrometry (TOF-SIMS) depth profiling was performed using an IONTOF TOF.SIMS 5 at the Colorado School of Mines. Negative secondary-ion spectra were acquired using a 30 keV $Bi_1^+$ primary analysis beam rastered over a 150 × 150 μm² area with 128 × 128 pixels. Depth profiling was performed in interlaced mode using a 1 keV $Cs^+$ sputter beam rastered over a 400 × 400 μm² area. $NO^-$ was used as the tracer for nitrogen because elemental $N^-$ exhibits a substantially lower secondary-ion yield in the $Cr_2O_3$ matrix. Profiles were collected from both an as-deposited reference and a 1000-hour-aged $Cr_2O_3$:N/$Ga_2O_3$ sample to evaluate the nitrogen distribution through the $Cr_2O_3$ layer and across the $Cr_2O_3/Ga_2O_3$ interface.

Specimens for scanning transmission electron microscopy (STEM) were prepared using a Tescan Solaris $Ga^+$ focused ion beam (FIB) using standard FIB lift out methods. STEM imaging and energy dispersive X-ray spectroscopy (EDS) data were acquired using a Thermo Fisher Scientific (TFS) Spectra 200 S/TEM operated at 200 kV with a convergence semi-angle of 24.2 mrad. EDS data were acquired on a TFS Super-X EDS detector.

## 3. Results:

### *3.1 Importance of thermodynamic stability*

Fabrication of p-n diodes for high temperature applications is an ongoing field of research; finding device architectures that can survive highly reducing environments at these temperatures has not been widely explored. Thermodynamic considerations of material stability becomes especially important when considering long-term operation of these devices. We have previously shown that a predominantly oxide-based architecture can reliably operate at high temperatures for long periods of time. Nickel oxide (NiO) is perhaps the mostly widely studied p-type material for $Ga_2O_3$ integration, due to its favorable band alignment with $Ga_2O_3$. However, there are several issues with its longevity, especially in the context of Ga2O3. Previous work has established the formation of a $NiGa_2O_4$ spinel structure that can form during deposition due to elevated temperatures (48).

As a proof of concept that demonstrates the need for thermodynamically stable materials and contacts, we fabricated and attempted to characterize a NiO/$Ga_2O_3$ sensor with a Pt catalyst. As shown in Fig.SX, the device failed catastrophically after several hundred hours. To rationalize the premature failure of the NiO/$Ga_2O_3$ device, Figure 1 shows both the Ellingham diagram for constituent oxide materials that would be considered in a typical $Ga_2O_3$ device stack (Figure 1, left panel); as well as requisite balance of hydrogen and humidity conditions that required for thermodynamic stability if that stack is operated as a hydrogen sensor (49,50) (right pane, dashed line). The NiO thermodynamics strongly suggests that it will almost immediately reduce to Ni metal in the presence of small amounts of hydrogen. We therefore hypothesize that the failed NiO device, which still showed a non-zero amount of activity in the presence of hydrogen, was reduced and subsequently reacted with Pt contact layer at 600˚C to form a Ni-Pt mixture or intermetallic. On the other hand, chromium oxide ($Cr_2O_3$), which is a relatively nascent p-type material candidate (44), is predicted to be stable in highly reducing environments. Hence, we further study doped $Cr_2O_3$ as a potential p-type material for $Ga_2O_3$ devices and their operation as high-temperature hydrogen sensors.

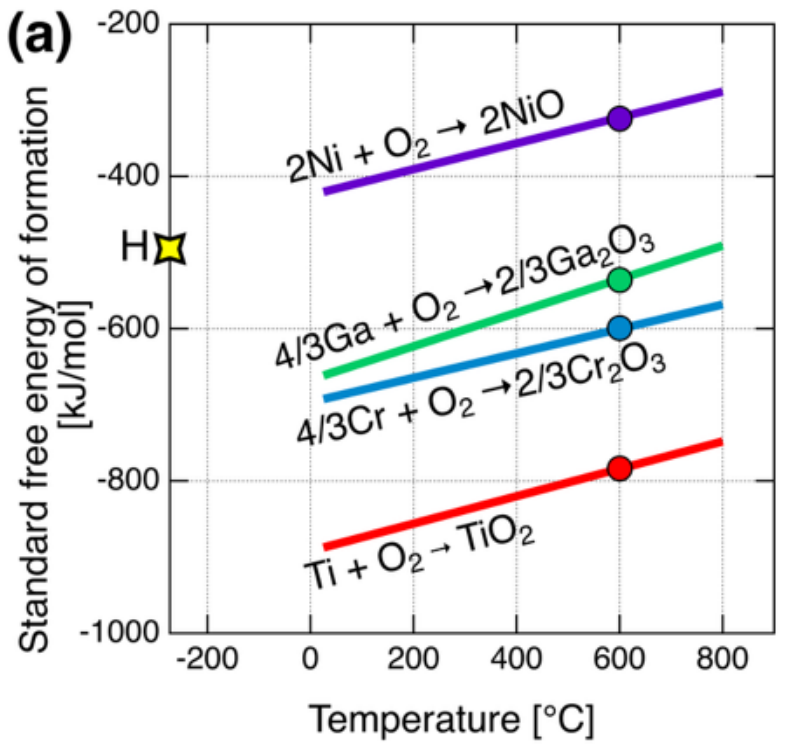

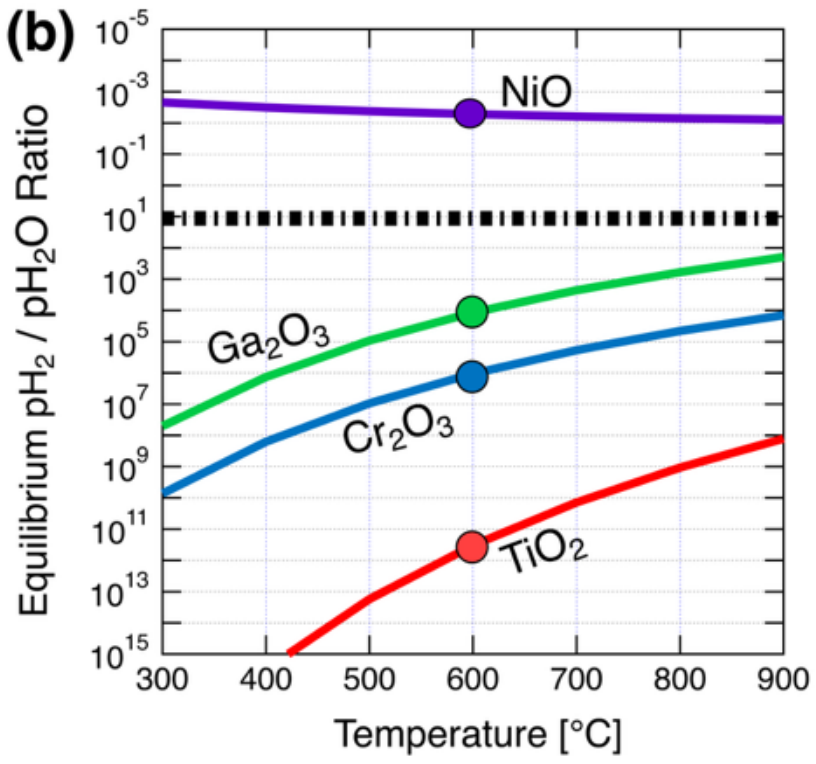

Figure 2 (Left) Plot of $\Delta G$ vs. $T$ for the metal oxides used in our hydrogen sensors, per mole of $O_2$. Reduction potential of hydrogen shown as a star. (Right) Plot of equilibrium ratio of $\frac{pH_2}{pH_2O}$ vs. T. The dashed line is an estimate of our experimental conditions. Materials above the dashed line are thermodynamically favored to be reduced in this environment, and those below are favored to be stable. The filled-in circles correspond to 600˚C operation temperature in both plots.

Prior to long-term testing, a "staircase" measurement was performed to characterize the $H_2$ sensing response of the $Cr_2O_3$:Mg device Figure 3 shows the resulting transient response, plotted as a function of both increasing and decreasing temperature and $H_2$ concentration, shows clear symmetry in the device's response to both variables. Notably, hydrogen sensitivity does not emerge until temperatures exceed 200°C. This can be explained through use of a combination of a Langmuir model for equilibrium hydrogen dissociation at the catalytic surface and a subsequent Helmholtz formulation to calculate the resulting collective dipole strength. While further studies are needed to fully understand this phenomenon, initial work suggests that the field is fully attenuated in the $Cr_2O_3$ layer until higher temperatures, at which point the potential barrier at the $Cr_2O_3/Ga_2O_3$ starts being affected.

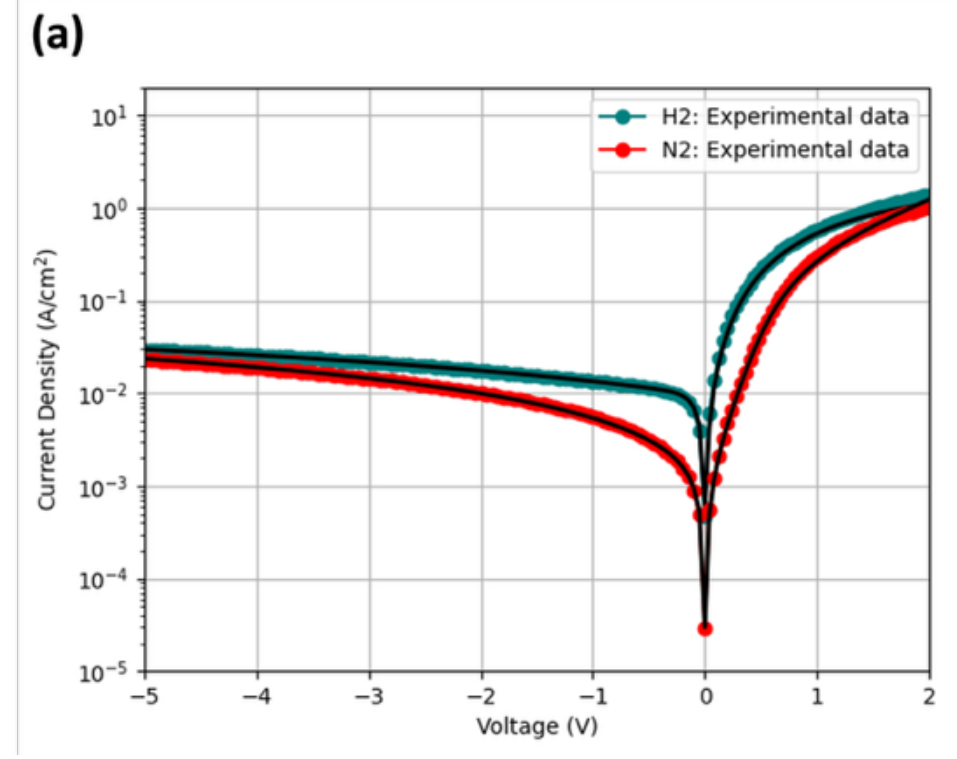

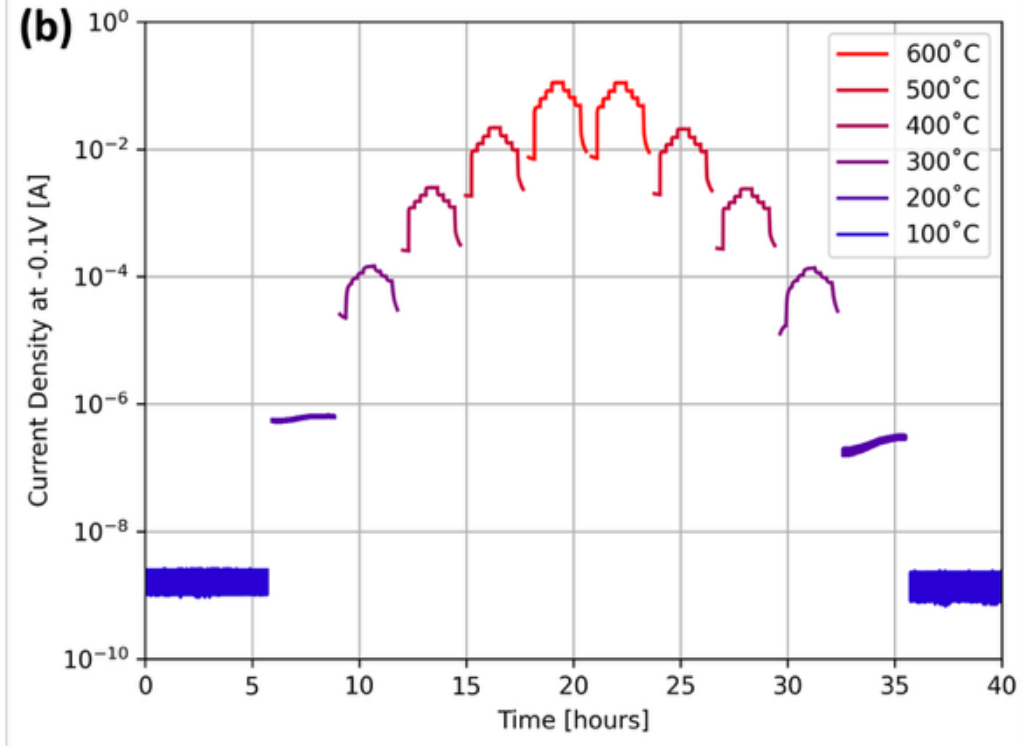

Figure 3: (a) Example current density – voltage (JV) measurement for a diode operating in nitrogen (red markers) and hydrogen (teal markers), showing the effects of hydrogen on the performance of the device at each voltage. The devices in this study showed a large differential in performance at -0.1 V; hence, this value was chosen as the "sensing voltage" for all long-term transient measurements. (b) Staircase measurement for a $Pt/Cr_2O_3$:Mg/$Ga_2O_3$ hydrogen sensor diode. The operating temperature of the diode was set between 100-600˚C in 100˚C steps. At each temperature, the concentration of hydrogen was set by blending 5% forming gas with a balance of $N_2$ to achieve concentrations of [0, 1250, 2500, 5000, 12500] ppm. The sensing current at -0.1 V was constantly recorded during the experiment. The device was allowed to equilibrate for 30 minutes at each temperature in $N_2$ to establish a baseline. This was followed by a 15-minute exposure to each gas concentration, both increasing and decreasing between minimum and maximum setpoints. IV curves were not collected during this experiment to reduce continuity mismatches between setpoint changes.

### *3.2 Long-term sensor measurements*

Figures 4 and Figure 5 present the transient current density response of each $Ga_2O_3/Cr_2O_3$ heterojunction sensor architecture, measured at -0.1 V, along with the resulting sensor signal which indicates how effectively the sensor operates in a dynamic gas environment. The representative JV curves of these sensors at the start and the end of the measurement period, with and without $H_2$ exposure, are also shown in Figures 4 and Figure 5 for the two heterojunction diodes, in comparison with a reference $Pt/Ga_2O_3$ Schottky barrier diode in Figure 6.

The $Cr_2O_3$:Mg device (Figures 4) displayed the most typical, expected degradation behavior among the architectures, consistent with previous observations. This data represents a sensor operating in an "always" on configuration, in which there are long periods of baseline / steady state operation (11 hours in $N_2$) , followed by shorter periods of sensing gas exposure (1 hour each of 500, 1000, and 1500 ppm $H_2$). Sensor signal dropped by over 60% of its initial value within the first 100–200 hours, followed by a much slower decline indicative of a "burn-in" period; this reduced degradation rate persisted until just after 1,000 hours, at which point the test was manually ended.

Figure 4c shows two sets of traces for these J-V curves measured using a switching unit: one set towards the beginning of the thermal soak, and one towards the end, for each device architecture. The $N_2$ traces are shown in black and 1,500 ppm $H_2$ curves in color, with earlier traces more transparent and later traces more opaque. These two sets of traces show that early in the soak, the enhanced hydrogen current response is distinct across all voltages, while over time this response becomes increasingly confined to forward bias and small reverse-bias voltages.

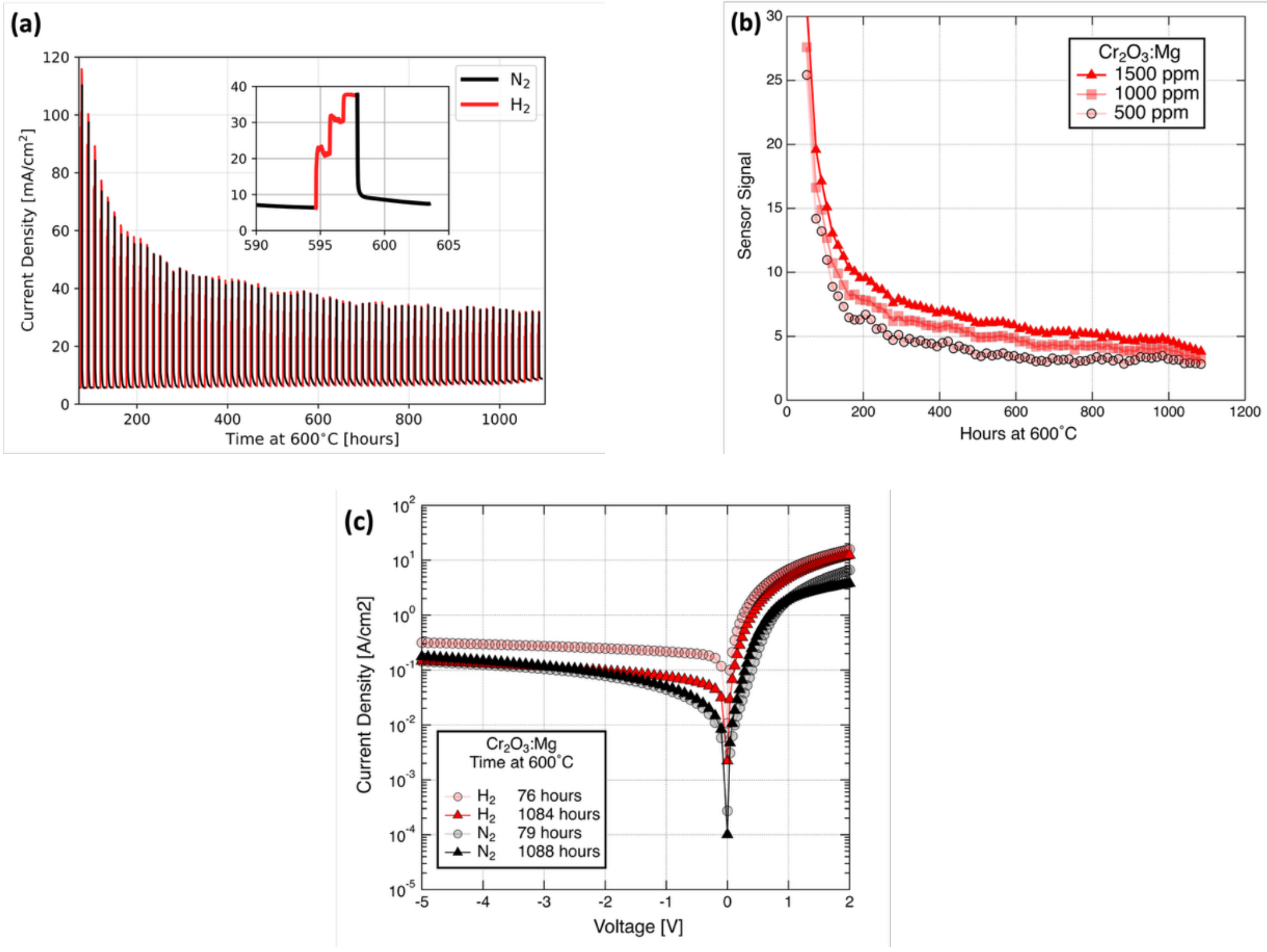

Figure 4: $Ga_2O_3$-based hydrogen sensors with $Cr_2O_3$:Mg, contact: (a) Transient response and (b) sensor signal of three different operating at -0.1 V, as a function of time at 600˚C. The inset of the (a) shows a zoomed-in view of this gas cycle, with distinct staircase steps corresponding to each of the hydrogen concentrations, followed by a return to baseline in $N_2$ at the end of the cycle. (c) JV Curves for $Cr_2O_3$:Mg device for $N_2$ (shown in black in all plots), as well as those collected in 1,500 ppm $H_2$ (shown in colored traces), at the start and end of the experiment.

A parallel set of experiments was conducted using a $Ga_2O_3$ sensor device with a nitrogen-doped p-type $Cr_2O_3$ layer, shown in Figure 5, evaluating J-V performance as a function of temperature from 50°C to 600°C alongside transient response measurements over more than 800 hours of operation. The $Cr_2O_3$:N device showed a lower overall performance than $Cr_2O_3$:Mg device, but a more desirable response profile, consistent with the JV measurements (Figure 5c). The $Cr_2O_3$:N device maintained a near-constant sensor signal across different hydrogen concentrations until its abrupt failure just after 800 hours. A minor discontinuity occurs at approximately 300 hours due to a lab power outage, after which testing resumed promptly and device performance normalized and stabilized over the subsequent 500 hours.

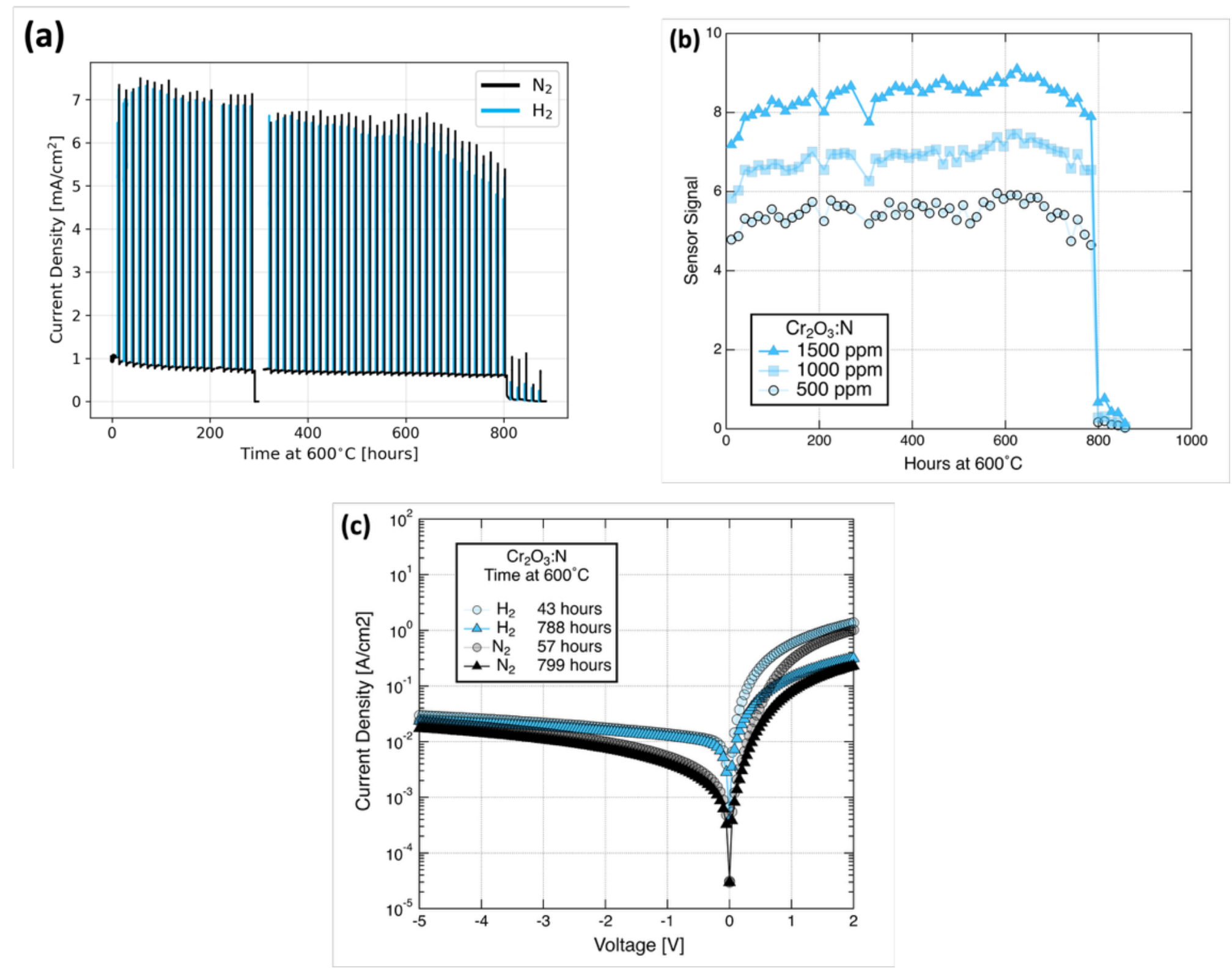

Figure 5: $Ga_2O_3$-based hydrogen sensors with $Cr_2O_3$:N layer: (a) Transient response and (b) sensor signal of three different operating at -0.1 V, as a function of time at 600˚C. Devices are operated continuously in this experiment, minus one instance of a power outage that required the system to be re-set to testing conditions. (c) JV Curves for $Cr_2O_3$:N device in $N_2$ (shown in black in all plots), as well as those collected in 1,500 ppm $H_2$ (shown in colored traces), at the start and end of the experiment.

Despite the decline in signal observed over time for $Cr_2O_3$:Mg device, and relatively lower signal for the $Cr_2O_3$:N device, both sensor architectures ($Cr_2O_3$:Mg and $Cr_2O_3$:N) remained capable of reliably detecting and clearly differentiating between three distinct, low hydrogen concentrations even after over a 800-1000 hours (over 1 month) of continuous operation at 600°C.

To isolate the role of the platinum catalyst layer from the broader device architecture, a Schottky diode with a 30 nm Pt layer was fabricated and tested at 600°C for 1,800 hours, as shown in Figure 6. The Pt Schottky diode exhibited the highest overall sensor performance, likely due to the absence of an attenuating p-type layer, though it also showed the greatest variability of the three architectures and was operated for the longest duration. The J-V evolution shows a stable $N_2$ response throughout testing, while the hydrogen response progressively increases toward higher leakage current. Device performance remains stable for the first 1,000 hours, exhibiting a break-in period during the initial 100–200 hours; beyond 1,000 hours, an accelerated degradation phase emerges near 1,100 hours, followed by a brief return to normal behavior around 1,250 hours. This degradation then resumes and grows exponentially until approximately 1,600 hours, at which point the device experienced catastrophic failure.

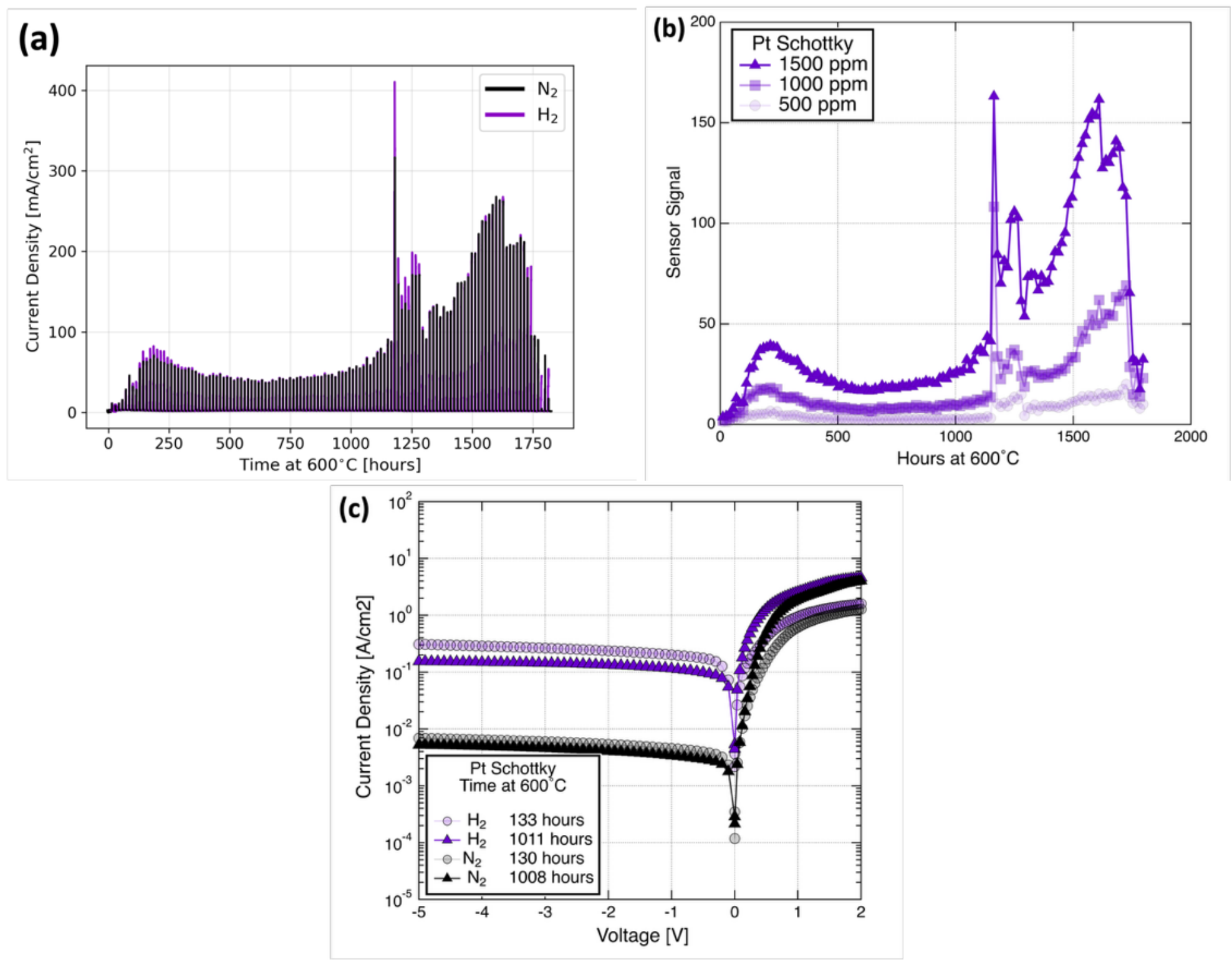

Figure 6: (a) Transient response and (b) sensor signal for the $Ga_2O_3$-based hydrogen sensors with Pt Schottky contacts operating at -0.1 V, as a function of time at 600˚C. (c) JV Curves for the Pt Schottky device for $N_2$ (shown in black in all plots), as well as those collected in 1,500 ppm $H_2$ (shown in colored traces), at the start and end of the experiment.

As shown by the sensor signal in Figure 6b, by the Pt Schottky device performance actually increased with time, especially in $N_2$. This is reflected mostly in the series resistance extracted from JV curves, shown for the 130 and 1000 hours in Figure 6c. Over the time, the series resistance which changes by almost an order of magnitude from 2.6 $\Omega$ cm$^2$ to a stable value of 0.28 $\Omega$ cm$^2$. This is potentially due to contact annealing of the Pt/$Ga_2O_3$ interface (51). It is worth noting that the majority of changes that occurred in the Pt Schottky device performance were during its exposure to $H_2$ – virtually all of the JV curves collected in $N_2$ were typical of an exceptional diode.

More information about the full evolution of the Pt Schottky device can be found in the Supplemental Information.

## 4. Discussion

To compare the sensor performance over time for all three device architectures, we calculated the sensor device sensitivity $\left(\frac{\Delta J_{H_2}}{\Delta C_{H_2}}\right)$, as shown in Figure 7a. For the $Cr_2O_3$:Mg device, sensitivity followed the same trend as the sensor signal, decreasing from an initial value greater than 30 µA/cm² to approximately 10 µA/cm² per ppm $H_2$. The $Cr_2O_3$:N device shows smaller but highly stable sensitivity, reflective of its lower sensor signal. On the other hand , the reference Pt Schottky device shows the highest but the least stable sensitivity over time

### *4.1 Electrical device analysis*

To analyze the sensor performance evolution from device model point of view, the J-V curves for each of the three devices were fit to a standard single diode equation using a differential evolution parameter extraction model (52,53), with reverse saturation current density ($J_0$), ideality factor (n), series resistance ($R_s$), and shunt resistance ($R_{sh}$) treated as fitting parameters. More information can be found in the Supplemental Information. These curves are fit to the diode equation:

$$J = J_0 \exp\left(\frac{V - JR_s}{nk_BT}\right) + \frac{V - JR_s}{R_{sh}}$$

Further, the parameter in the diode equation that describes leakage current, $J_0$, was used to calculate the barrier height that exists in the diode. While leakage current has previously been ascribed to Poole-Frenkel emission (44), and Schottky diodes are traditionally described using thermionic emission (39–41), the *change* in barrier height reduces to the same mathematical form, regardless of which of these two frameworks is being considered:

$$\Delta\phi_B = -k_BT \ln\left(\frac{J_{0,H_2}}{J_{0,N_2}}\right)$$

Hence, through use of just the extracted reverse leakage current density parameter and by treating the barrier as generic instead of associating it with any particular mechanism, the effect of the presence of hydrogen can be quantified and directly compared between the three different devices.

To quantify the mechanism driving the changing sensor signal and the resulting sensitivity (Figure 6) in three sensor devices, we extracted the series resistance from all the measured JV curves. The series resistance in the $Cr_2O_3$:N device showed instead increased at an approximately constant rate from 1 $\Omega\, cm^2$ to 4 $\Omega\, cm^2$ before its catastrophic failure. The changes in the $Cr_2O_3$:Mg fits are the opposite of this – there is a strong dichotomy between the $H_2$ and $N_2$ curves, with the largest changes coming in the $N_2$ curves. Series resistance increased from an initial value of 0.15 $\Omega\, cm^2$ to a maximum of 0.5 $\Omega\, cm^2$. This behavior is likely linked to the growth of an MgO layer at the $Cr_2O_3/Ga_2O_3$ interface (44), which would further attenuate the proton-induced dipole effect at the catalyst layer.

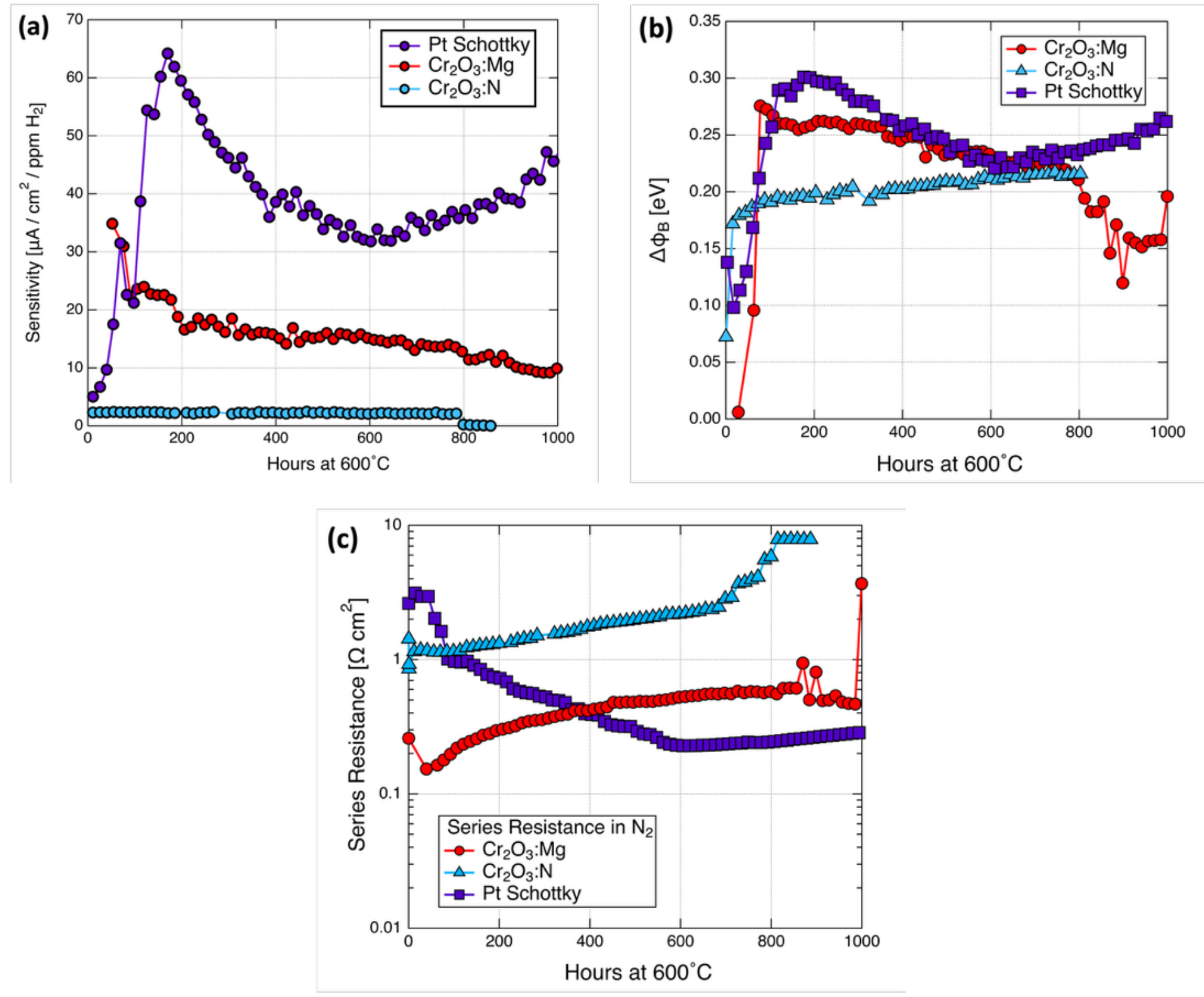

Figure 7: (a) Calculated sensitivity; (b), absolute change in barrier height as a function of time, calculated from the extracted leakage current density; and (c) extracted series resistance from $N_2$ IV curves for all three tested devices.

On the other hand, the operation of the electronic heterojunction itself, as characterized by the change in the barrier high, was similar for all three device architectures indicating the same detection mechanisms. The barrier height also remained relatively unchanged over the long periods of time. The extracted $J_0$ values were used with the thermionic emission expression to solve for barrier height, assuming a Richardson coefficient ($A^*$) of 41 A/(cm²·K) (54,55), with all remaining variables taken as known or measured quantities. The results of the effective barrier calculation using $J_0$ values extracted from JV curves is shown in Figure 7b for all three traces, with only the first 1,000 hours of the Pt Schottky being considered to appropriately window the x-axis. This plot represents the absolute value of the change that's occurring between gas cycles, with the physical phenomenon corresponding to a decrease in effective barrier height in the hydrogen environment compared to the inert environment. In all cases, the change in effective barrier height in the diode is significant (> 0.18 eV post break-in), and as expected, the trends in performance are consistent with what has been previously described, with the Pt Schottky experiencing the largest overall change, followed by the $Cr_2O_3$:Mg device with a similar but slightly diminished performance shape. The smallest but most consistent changes to the barrier were seen in the $Cr_2O_3$:N device; likewise, the break-in period was completed within the first few gas cycles.

It should be noted that there is a possibility of misattribution in any parameter extraction implementation, especially in forward bias between the ideality factor and the series resistance. It is therefore more instructive to look at the evolutions internal to each device instead of treating cross-device parameter comparisons as definitive. More objective metrics, such as rectification ratios or parameters that cannot be as readily misattributed (such as reverse leakage current density), can readily be used to compare devices. Table 1 shows the rectification ratio of each diode in both $N_2$ and in 1,500 ppm $H_2$, measured at $\pm 2$ V, with the "early" and "late" demarcations corresponding to the times shown in Figure 6. The colors in the "Late" columns indicate an improvement (green, bold) or worsening (red, italics) of performance relative to the "Early" columns. Overall these numbers extracted from the JV curves directly appear consistent with the fitting and modeling the JV curves, increasing confidence in our analysis.

Table 1: Approximate rectification ratio (rounded to integer values) for each of the device architectures at $\pm 2$ V. "Early" and "Late" correspond to the timestamps shown for the JV curves in Figure 6.

| | $N_2$ Early | $N_2$ Late | 1500 ppm $H_2$ Early | 1500 ppm $H_2$ Late |
|---|---|---|---|---|
| $Cr_2O_3$:Mg | 85 | *44* | 64 | **128** |
| $Cr_2O_3$:N | 102 | *31* | 77 | *20* |
| Pt Schottky | 238 | **995** | 7 | **34** |

#### *4.2 Post-Mortem Analysis*

A pn diode with a thicker $Cr_2O_3:N_2$ layer was grown and analyzed via TOF-SIMS, using $NO^-$ as the tracer species due to the significantly lower secondary ion yield of elemental $N^-$ in this oxide matrix (Figure 8). The as-deposited sample showed good nitrogen incorporation within the $Cr_2O_3$ layer with virtually no spill-over into the $Ga_2O_3$. A 1000-hour tested sample, which had a thinner $Cr_2O_3$ layer, was also analyzed by TOF-SIMS and showed nitrogen incorporation at a lower level than the pristine sample but similarly exhibited an abrupt compositional edge at the $Cr_2O_3/Ga_2O_3$ interface with no detectable nitrogen incorporation into the $Ga_2O_3$ substrate. These results support the hypothesis that using alternative dopants in place of magnesium can help produce stable devices capable of long-term operation under extreme conditions.

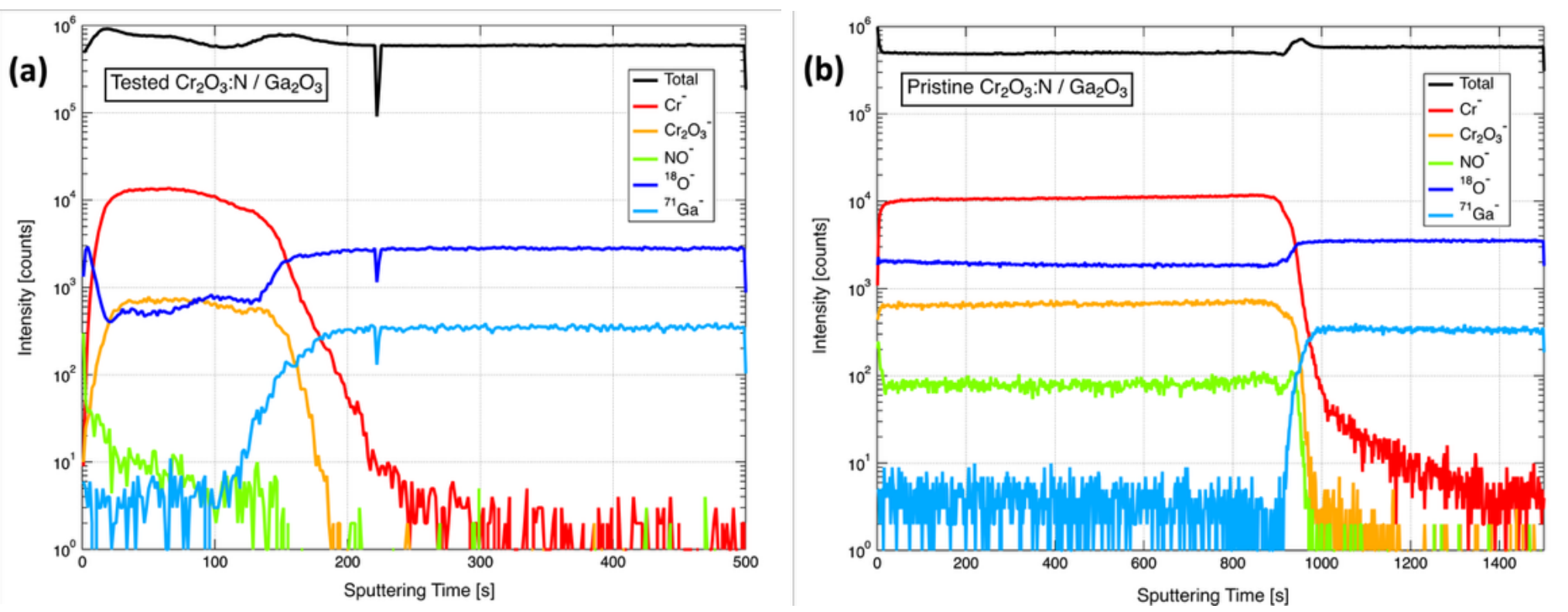

Figure 8 SIMS analysis of 1000-hour tested $Cr_2O_3$:N sample (left) and an as-fabricated reference sample. The reference sample was grown with the same nominal growth conditions but more pulses to emphasize the nitrogen incorporation.

A Pt Schottky diode operated for 1800 hours was analyzed via TEM alongside a pristine, as-deposited reference sample (same e-beam deposition parameters) to assess morphological changes at the Pt/$Ga_2O_3$ interface (Figure 9). In the pristine sample, the Pt/W protective layer interface was difficult to resolve due to similar Z-contrast (and possible surface damage from an overly aggressive W deposition), while high-resolution imaging revealed a polycrystalline Pt film with small grains (~5–10 nm) and a $Ga_2O_3$ interface that was not atomically flat. The substrate/Pt interface appeared slightly uneven, with apparent nanoscale intermixing attributed to a projection artifact from lamella thickness and substrate waviness rather than true interfacial reaction, and no interface reconstruction was observed. In contrast, the operated device showed substantial Pt grain growth, with grains coarsening to the tens-of-nanometers range, along with the formation of pores in the film and Pt "patches" on the surface that were likely observable in prior SEM imaging. HAADF imaging showed dark void regions at the film/substrate interface, and since the oxygen signal was higher within these voids than in the Pt film itself, the bulk Pt does not appear to have oxidized — consistent with the highly reducing operating conditions. Film thickness remained consistent with the recipe at 30–40 nm. Overall, device operation drove significant Pt film restructuring through grain growth and microvoid formation at the film/substrate interface, while the bulk Pt remained metallic and the film thickness was preserved.

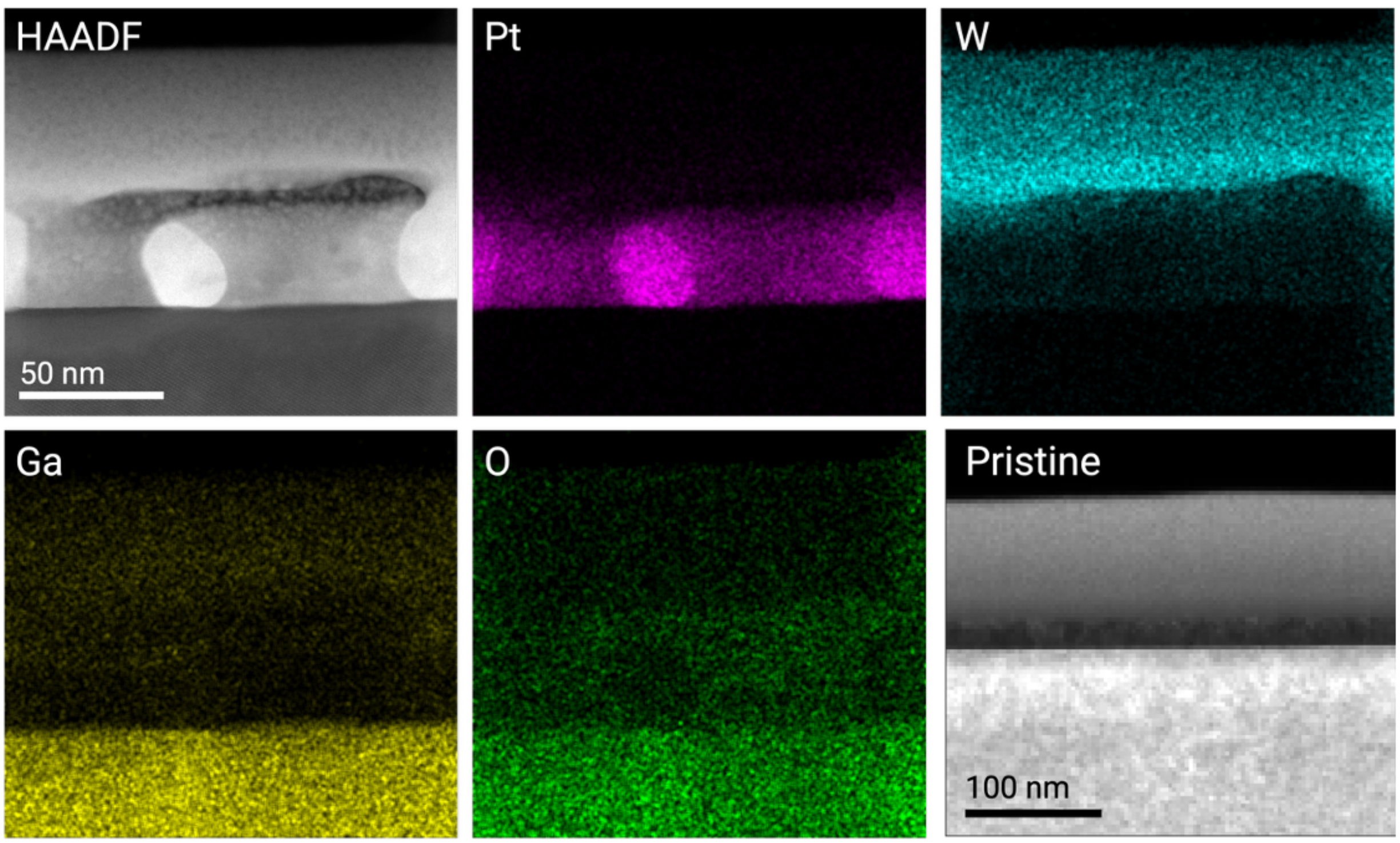

Figure 9: TEM analysis of Pt Schottky devices. An as-fabricated "pristine" device was included as a reference. The top two panes show 4D-STEM images of cross sections of the Pt/$Ga_2O_3$ device. The pristine device shows an abrupt and continuous interface between the two layers. The device operated for 1800 hours also shows this abrupt interface, but micro voids have formed and interspersed within the Pt layer. EDS (center and bottom row panes) confirms the agglomeration of Pt, likely resulting in these micro voids. It does not appear that the bulk of the Pt film has oxidized.

**5. Summary and Conclusions:**

This work presents a comprehensive investigation into the long-term reliability and sensing performance of $Ga_2O_3$-based hydrogen sensors, employing both Pt Schottky and $Cr_2O_3$/$Ga_2O_3$ p-n heterojunction diode architectures. Continuous operation over extended thermal soaks—exceeding 1,000 hours for the p-n devices and nearly 1,800 hours for the Pt Schottky diodes at 600°C—demonstrated that these sensors are capable of reliably detecting and differentiating low concentrations of hydrogen (500–1,500 ppm) over many weeks of continuous operation, despite a gradual decline in sensor signal and sensitivity over time. Analysis of the transient and J-V response revealed a consistent multi-stage degradation pattern across device architectures: an initial break-in period within the first 100–200 hours, followed by a period of relative stability, and ultimately either a gradual or catastrophic decline in performance at longer timescales (1,200–1,800 hours for Pt Schottky devices, ~800 hours for the p-n devices).

Diode parameter extraction using thermionic emission and Lambert W-based fitting models confirmed that hydrogen exposure lowers the effective barrier height at the metal-semiconductor and heterojunction interfaces, consistent with a proton-induced dipole mechanism that thins the barrier and facilitates enhanced electron tunneling. Microstructural analysis (TEM) of aged Pt Schottky devices linked long-term degradation to physical changes at the Pt/$Ga_2O_3$ interface, including significant Pt grain growth and micro void formation, while confirming that the bulk Pt film remains metallic rather than oxidizing under the reducing sensing environment. Complementary TOF-SIMS analysis of the $Cr_2O_3$:N2 p-n devices showed that nitrogen doping remains well-confined within the $Cr_2O_3$ layer with minimal diffusion into the $Ga_2O_3$ substrate, even after extended operation, supporting the use of nitrogen as a stable alternative to magnesium doping and helping to explain the improved stability of these devices relative to Mg-doped counterparts, where prior work identified interfacial MgO formation as a key degradation pathway.

Temperature- and concentration-dependent staircase measurements further established that hydrogen sensitivity in the $Cr_2O_3$-based devices is thermally activated, with no measurable response below approximately 200°C—likely governed by the activation energy for hydrogen dissociation, transport limitations through the $Cr_2O_3$ layer, or temperature-dependent Debye screening effects. Taken together, these results indicate that both device architectures are viable candidates for high-temperature, long-duration hydrogen sensing in extreme environments, with degradation driven primarily by interfacial and microstructural evolution rather than bulk oxidation or catastrophic chemical failure. Future work, including analogous temperature-dependent testing of the Pt Schottky architecture and continued refinement of alternative doping strategies, will help further decouple the roles of the catalytic metal layer, the semiconductor heterojunction, and interfacial chemistry in determining sensor lifetime and low-temperature operability.

**Acknowledgements**

This work was authored by the National Laboratory of the Rockies for the U.S. Department of Energy (DOE), operated under Contract No. DE-AC36-08GO28308. Funding is provided primarily by the Office of Critical Minerals and Energy Innovation (CMEI) Advanced Materials & Manufacturing Technologies suboffice (AMMTO). The views expressed in the article do not necessarily represent the views of the DOE or the U.S. Government. This material makes use of the TOF-SIMS system at the Colorado School of Mines, which was supported by the National Science Foundation under Grant No. 1726898.

**Credit:**

**WC**: Device fabrication (substrate etching, metallization), experimental design, data acquisition and analysis, manuscript conceptualization, writing and editing. **KE**: PLD: Device fabrication for $Cr_2O_3$:Mg and $Cr_2O_3$:N samples. **A.Sacchi**: PLD: Device Fabrication for $Cr_2O_3$:N samples for SIMS analysis. **MS**: Microscopy and analysis. **MW**: SIMS analysis. **A.Staerz**: Assistance in data analysis / discussion. **RO**: Assistance in data analysis / discussion. **AZ**: Project lead, manuscript conceptualization.

Supplemental Information

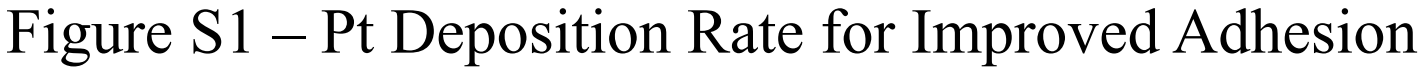

Figure S1 – Pt Deposition Rate for Improved Adhesion

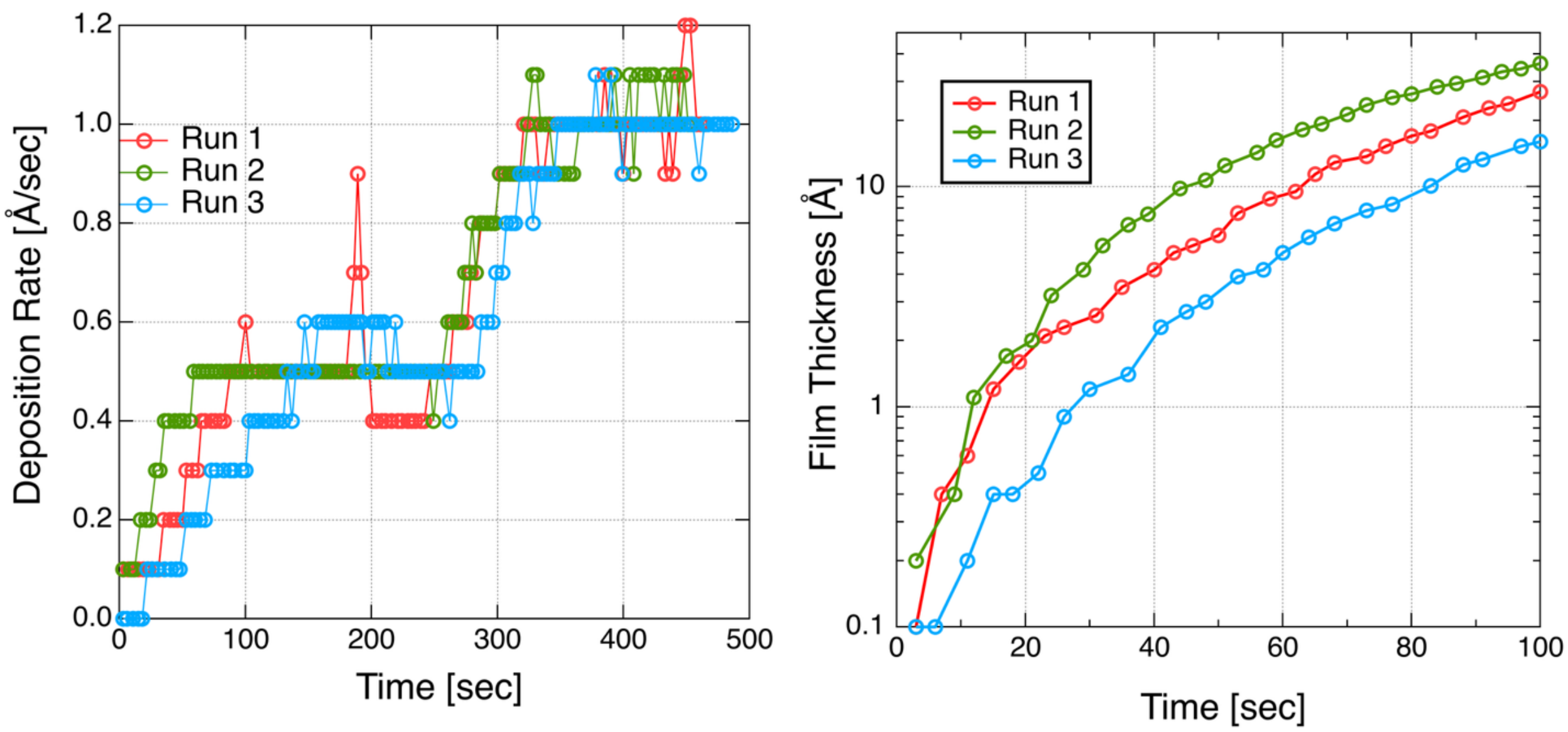

A "slow" platinum deposition recipe was developed in the Temescal FC2000 e-beam evaporation system, in which the first 10 nm of the Pt film was deposited at 0.5 Å/s and the remaining 20 nm was deposited at 1 Å/s. The left plot shows deposition rate as a function of time for the first 500 seconds, illustrating a very slow ramp rate during the initial few monolayers of Pt deposition, while the right plot shows film thickness on a log scale versus time for the first 10 Å; together, these plots demonstrate the intentionally slow initial deposition of platinum onto the Ga2O3 substrate, which promotes better adhesion between the noble metal and the polished substrate. Film thickness was monitored using a quartz crystal monitor (QCM), and slight run-to-run variations can be attributed to QCM status, deposition base pressure, and the condition of the Pt target.

Figure S2 – Educational IV Curves, etc.

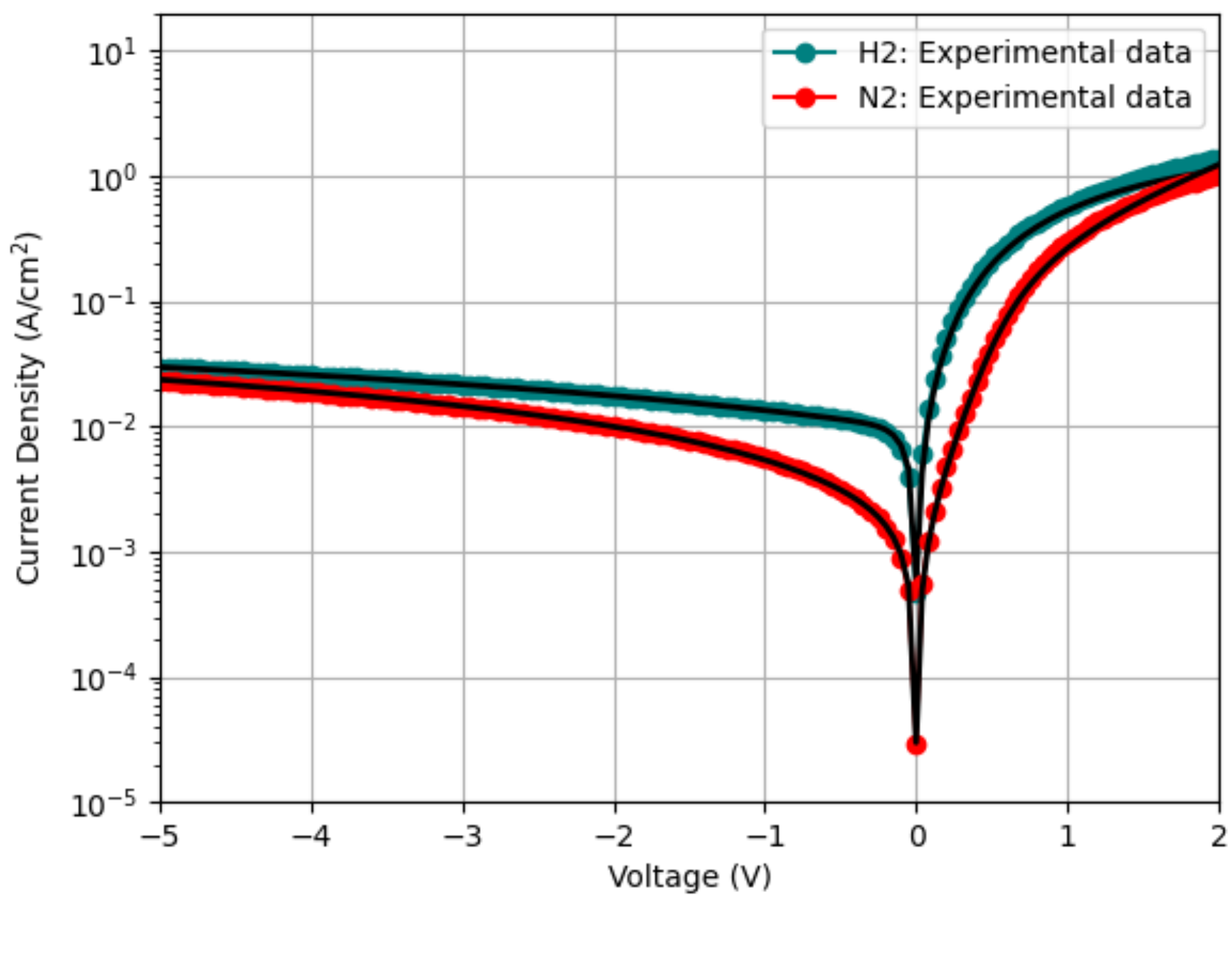

$$J = J_0 \exp\left(\frac{V - JR_s}{nk_BT}\right) + \frac{V - JR_s}{R_{sh}}$$

The Lambert W implementation (black lines) fit each J-V curve with a mean absolute percent error of less than 5% and was primarily used to extract the reverse saturation current density, from which the barrier height change between $N_2$ and 1,500 ppm $H_2$ conditions was calculated. The difference between the $N_2$ and $H_2$ barrier heights ($\Delta\phi_B$) is plotted in Figure 7b, demonstrating that hydrogen exposure effectively reduces the potential barrier between the p-type and n-type layers, allowing current to flow more easily through the diode. This plot reveals several additional trends: $\Delta\phi_B$ decreases steadily at first and then more rapidly over the final 100–200 hours of soak time, and since the $N_2$ barrier height remains roughly constant at 1.6–1.7 eV throughout, this decline indicates that hydrogen's ability to reduce the barrier height diminishes with time. There is also substantial device-to-device variance, with a large difference in $\Delta\phi_B$ between D1/D3 and D2 that meaningfully affects device performance; fully characterizing this statistical spread would require a specialized movable-probe-tip instrument (Callahan et al., 2023). Finally, the evolution of $\Delta\phi_B$ over time points to an underlying morphological change in the device—prior work showed that Mg dopant in the $Cr_2O_3$ layer migrates to the $Cr_2O_3$ /$Ga_2O_3$ interface and oxidizes into a thin MgO layer, increasing resistance under inert conditions (Callahan et al., 2024), and a similar interfacial change is likely occurring here and affecting the extracted barrier height.

Figure S3 – Evolution of Pt JV Curves with Time, Through Failure.

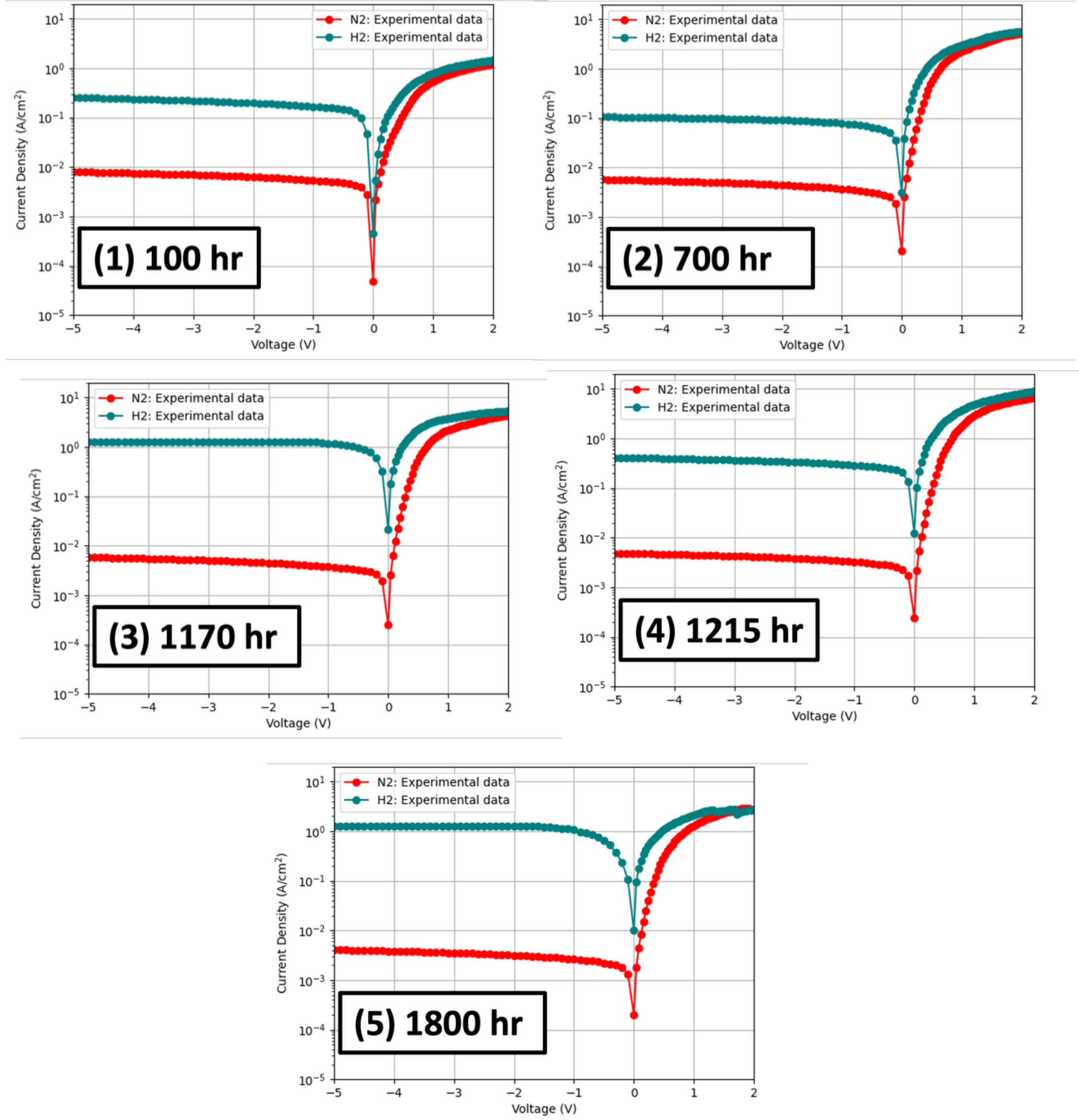

The figure presents a time-series evolution of I-V characteristics under both N2 and 1,500 ppm H2 exposure at five representative timestamps, encompassing over 125 gas cycles across nearly 1,800 hours of continuous operation at 600°C for a Ga2O3-based hydrogen sensor. At 100 hours, the device approaches break-in and exhibits improved forward-bias behavior relative to initial conditions, with the most significant change observed in reverse bias, where hydrogen exposure yields a current increase exceeding one order of magnitude. By 700 hours, device performance continues to improve, demonstrating excellent rectification and hydrogen sensitivity. At approximately 1,169 hours, an anomalous spike is observed, with the reverse-bias response becoming fully saturated and reaching the source meter's compliance limit, resulting in behavior that approximates a resistive element under these conditions. Approximately 50 hours later, at 1,213 hours, the hydrogen response recovers to a level slightly above that observed at 700 hours; across these first four timestamps, forward-bias behavior remains largely unaffected, and the N2 baseline curve remains nearly consistent following break-in. Finally, beyond 1,700 hours, the device exhibits failure in both forward and reverse bias: the reverse-bias current again reaches the compliance limit of the source meter, while the forward-bias response displays nonlinear behavior consistent with poor electrical contact. Notably, despite this degradation, the N2 baseline response remains highly stable throughout.

Figure S4 – Logistic Fit of Sensor Performance.

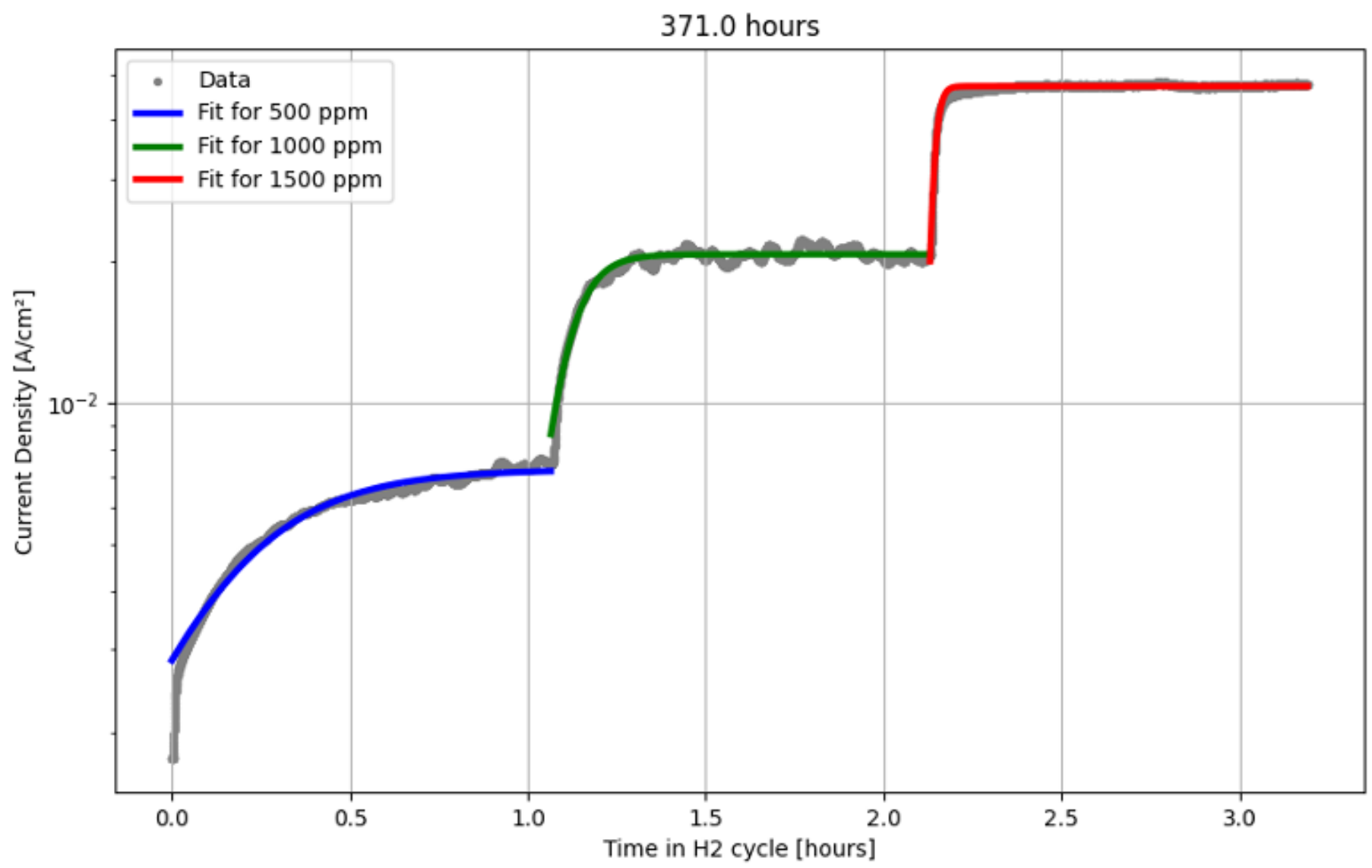

To characterize the transient behavior of the sensors at varying hydrogen concentrations, the current density at -0.1 V was recorded during the hydrogen exposure staircase steps at 500, 1,000, and 1,500 ppm, with each concentration step fit to a logistic growth function:

$$f(t) = \frac{L}{1 + \exp(-k(t - t_0))}$$

The primary parameter of interest, $L$, represents the steady-state plateau value; fitting this parameter provides a more rigorous approach than eyeballing or averaging the raw data, and the resulting $L$ values were used to calculate sensor response and, subsequently, sensitivity. The plot shows a strong fit across all concentrations, with the sharpness of the logistic growth curve becoming more pronounced as concentration increases. At lower concentrations, the response appears comparatively sluggish, suggesting the sensor was operating near the onset of its sensitivity threshold

Figure S5: Attempted Transient Characterization of Pt/NiO/$Ga_2O_3$ Device.

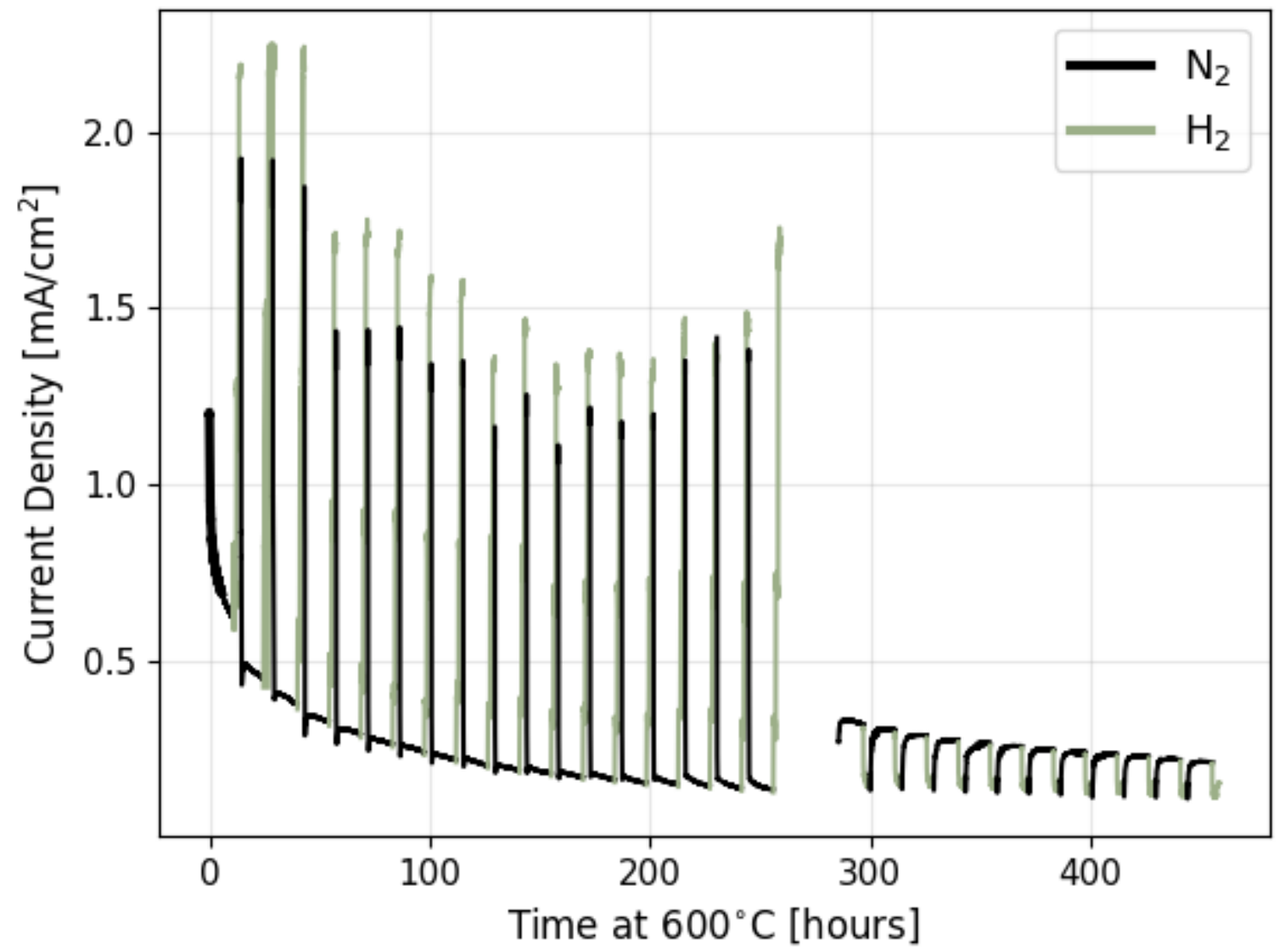

Figure S5 shows the transient performance of a Pt/NiO/$Ga_2O_3$ heterojunction diode operated at -0.1V. Device performance is poor, with a sensor response less than 5 for all concentrations of hydrogen. Based on thermodynamics and the high operating temperature of the device, it's likely that the NiO layer of the device was reduced to Ni metal during the first hydrogen gas cycle, leaving a Pt/Ni catalyst layer in contact with the $Ga_2O_3$ substrate. Previous reports have shown that layering Pt and Ni metals will still yield hydrogen dissociation activity (56), which explains the retained hydrogen response. However, this demonstration shows that stable alternatives to NiO must be considered for broad-ranging applications of $Ga_2O_3$ devices.